\documentclass[twoside,twocolumn,9pt]{article}
\usepackage{extsizes}
\usepackage[super,sort&compress,comma]{natbib} 
\usepackage[version=3]{mhchem}
\usepackage[left=1.5cm, right=1.5cm, top=1.785cm, bottom=2.0cm]{geometry}
\usepackage{balance}
\usepackage{mathptmx}
\usepackage{sectsty}
\usepackage{graphicx} 
\usepackage{lastpage}
\usepackage[format=plain,justification=justified,singlelinecheck=false,font={stretch=1.125,small,sf},labelfont=bf,labelsep=space]{caption}
\usepackage{float}
\usepackage{fancyhdr}
\usepackage{fnpos}
\usepackage[english]{babel}
\addto{\captionsenglish}{%
  
}
\usepackage{array}
\usepackage{droidsans}
\usepackage{charter}
\usepackage[T1]{fontenc}
\usepackage[usenames,dvipsnames]{xcolor}
\usepackage{setspace}
\usepackage[compact]{titlesec}
\usepackage{hyperref}
\usepackage{gensymb}
\usepackage{wasysym}
\usepackage{epstopdf}
\usepackage{lipsum}
\usepackage{graphicx}
\definecolor{cream}{RGB}{222,217,201}

\begin{document}

\pagestyle{fancy}
\thispagestyle{plain}
\fancypagestyle{plain}{
\renewcommand{\headrulewidth}{0pt}
}

\makeFNbottom
\makeatletter
\renewcommand\LARGE{\@setfontsize\LARGE{15pt}{17}}
\renewcommand\Large{\@setfontsize\Large{12pt}{14}}
\renewcommand\large{\@setfontsize\large{10pt}{12}}
\renewcommand\footnotesize{\@setfontsize\footnotesize{7pt}{10}}
\makeatother

\renewcommand{\thefootnote}{\fnsymbol{footnote}}
\renewcommand\footnoterule{\vspace*{1pt}%
\color{cream}\hrule width 3.5in height 0.4pt \color{black}\vspace*{5pt}} 
\setcounter{secnumdepth}{5}

\makeatletter 
\renewcommand\@biblabel[1]{#1}            
\renewcommand\@makefntext[1]%
{\noindent\makebox[0pt][r]{\@thefnmark\,}#1}
\makeatother 
\renewcommand{\figurename}{\small{Fig.}~}
\sectionfont{\sffamily\Large}
\subsectionfont{\normalsize}
\subsubsectionfont{\bf}
\setstretch{1.125} 
\setlength{\skip\footins}{0.8cm}
\setlength{\footnotesep}{0.25cm}
\setlength{\jot}{10pt}
\titlespacing*{\section}{0pt}{4pt}{4pt}
\titlespacing*{\subsection}{0pt}{15pt}{1pt}

\fancyfoot{}
\fancyfoot[LO,RE]{\vspace{-7.1pt}}
\fancyfoot[CO]{\vspace{-7.1pt}\hspace{11.9cm}}
\fancyfoot[CE]{\vspace{-7.2pt}\hspace{-13.2cm}}
\fancyfoot[RO]{\footnotesize{\sffamily{1--\pageref{LastPage} ~\textbar  \hspace{2pt}\thepage}}}
\fancyfoot[LE]{\footnotesize{\sffamily{\thepage~\textbar\hspace{4.65cm} 1--\pageref{LastPage}}}}
\fancyhead{}
\renewcommand{\headrulewidth}{0pt} 
\renewcommand{\footrulewidth}{0pt}
\setlength{\arrayrulewidth}{1pt}
\setlength{\columnsep}{6.5mm}
\setlength\bibsep{1pt}

\makeatletter 
\newlength{\figrulesep} 
\setlength{\figrulesep}{0.5\textfloatsep} 

\newcommand{\topfigrule}{\vspace*{-1pt}%
\noindent{\color{cream}\rule[-\figrulesep]{\columnwidth}{1.5pt}} }

\newcommand{\botfigrule}{\vspace*{-2pt}%
\noindent{\color{cream}\rule[\figrulesep]{\columnwidth}{1.5pt}} }

\newcommand{\dblfigrule}{\vspace*{-1pt}%
\noindent{\color{cream}\rule[-\figrulesep]{\textwidth}{1.5pt}} }

\makeatother

\twocolumn[
  \begin{@twocolumnfalse}
{
\includegraphics[width=18.5cm]{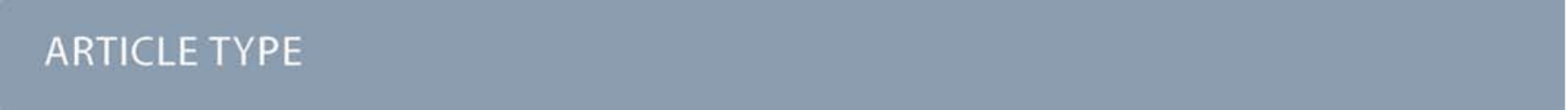}}\par
\vspace{1em}
\sffamily
\begin{tabular}{m{4.5cm} p{13.5cm} }

&
\noindent\LARGE{\textbf{Polytypism and Superconductivity in the NbS$_2$ System}} \\
\vspace{0.3cm} & \vspace{0.3cm} \\

 & \noindent\large{Catherine Witteveen,\textit{$^{a,b}$} Karolina Gornicka,\textit{$^{c,d}$}  Johan Chang,\textit{$^{b}$} Martin M\aa nsson,\textit{$^{e}$} Tomasz Klimczuk,\textit{$^{c,d}$} Fabian O. von Rohr,\textit{$^{a,b}$}$^{\ast}$} \\%

\includegraphics{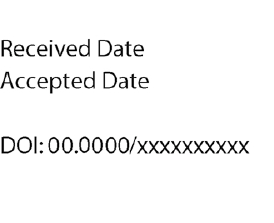} & \noindent\normalsize{

We report on the phase formation and the superconducting properties in the \ce{NbS2} system. Specifically, we have performed a series of standardized solid-state syntheses in this system, which allow us to establish a comprehensive synthesis map for the formation of the two polytypes 2H-\ce{NbS2} and 3R-\ce{NbS2}, respectively. We show that the identification of two polytypes by means of X-ray diffraction is not always unambiguous. Our physical property measurements on a phase-pure sample of 3R-\ce{NbS2}, on a phase-pure sample of 2H-\ce{NbS2}, and a mixed phase sample confirm earlier reports that 2H-\ce{NbS2} is a bulk superconductor and that 3R-\ce{NbS2} is not a superconductor above $T =$ 1.75 K. Our results clearly show that specific heat measurements, as true bulk measurements, are crucial for the identification of superconducting materials in this and related systems. Our results indicate that for the investigation of van-der-Waals materials great care has to be taken on choosing the synthesis conditions for obtaining phase pure samples.
}
 \\

\end{tabular}

 \end{@twocolumnfalse} \vspace{0.6cm}

  ]

\renewcommand*\rmdefault{bch}\normalfont\upshape
\rmfamily
\section*{}
\vspace{-1cm}


\footnotetext{\textit{$^{a}$~ Department of Chemistry, University of Zürich, Winterthurerstr. 190, 8057 Zürich, Switzerland }}
\footnotetext{\textit{$^{b}$~Department of Physics, University of Zürich, Winterthurerstr. 190, 8057
Zürich, Switzerland }}
\footnotetext{\textit{$^{c}$~Department of Solid State Physics, Gdansk University of Technology,80-233 Gdansk, Poland }}
\footnotetext{\textit{$^{d}$~Advanced Materials Centre, Gdansk University of Technology,
ul. Narutowicza 11/12, 80-233 Gdansk, Poland}}
\footnotetext{\textit{$^{e}$~Department of Applied Physics, KTH Royal Institute of Technology, Roslagstullsbacken 21, SE-106 91 Stockholm, Sweden }}
\footnotetext{$^{\ast}$To whom correspondence may be addressed.}



\section{Introduction}
Layered transition-metal dichalcogenides (TMDs) with the general formula \ce{MX2} (with M = group IV transition metal, group V transition metal, or Re and X = S, Se, or Te) have been of great interest due to their rich electronic properties, in combination with the opportunity to exfoliate them down to the monolayer.\cite{Coleman2011a,Lotsch2015b,Ryu2018} Recently, TMD superconductors have been identified as ideal model systems for investigating superconductivity in the two-dimensional limit.\cite{Xi2015a,Ryu2018,Devarakonda2019} Especially noteworthy is the realization of an intrinsic monolayer superconductor - i.e. the occurrence of superconductivity without the need of a specialized substrate - in monolayers of \ce{NbSe2}.\cite{Xi2015a,Wang2017,Xing2017} This, together with the observation of unconventional superconducting properties in TMD bulk superconductors - e.g. the linear scaling of the superfluid density in \ce{NbSe2}, or the observation of time-reversal symmetry breaking in superconducting 4H$_b$-\ce{TaS2} - hint towards new opportunities to potentially discover unconventional or even topological superconductivity using van der Waals materials or heterostructures thereof.\cite{VonRohr2019a,Ribak2020}

Generally, TMDs are subjected to structural polymorphism and polytypism, meaning there are several different phases with different crystal structures for the same chemical composition. In TMDs different polymorphs  occur by the changing coordination of the chalcogen to the metal atom, and the different polytypes by changing stacking sequences of the \ce{MX2} monolayers. Polytypism is occurring in layered materials, namely when the geometry of a repeating structural layer is maintained but the layer-stacking sequence of the overall crystal structure can be varied.\cite{Brown1965,Luo2015} Naturally, the crystal structure of a material defines its physical properties, hence these can vary drastically among different TMD polymorphs. For example, while the 1T'-\ce{MoTe2} polymorph is a Weyl semimetal and superconductor, the 2H-\ce{MoTe2} polymorph is a semiconductor with an indirect bandgap of $E_{\rm gap} =$ 1.0 eV.\cite{Qi2016a,Keum2015,Guguchia2017,Guguchia2018,Santos2020} However, physical properties of different polytypes are usually similar, since the changes in layer stacking impact the properties of the whole material in a less pronounced fashion.\cite{Choyke2013} For example, the 2H-\ce{NbSe2} and 4H-\ce{NbSe2} polytypes of \ce{NbSe2} are both superconductors with similar critical temperatures of $T_{\rm c} =$ 7.2 K and 6.5 K, and both show charge-density wave ordering at $T_{\rm CDW} = $ 35 K and 42 K, respectively.\cite{Naik2011,VonRohr2019a}

Among the metallic TMDs, the niobium disulfide system stands out, as none of its polymorphs have been reported to display charge-density-wave ordering.\cite{Leroux2012,Leroux2018} In this system, there are three polymorphs known. The stable 3R-\ce{NbS2} and 2H-\ce{NbS2} polytypes have been reported as polycrystalline and single crystalline materials, while the metastable 1T-\ce{NbS2} polymorph has been stabilized in thin film form.\cite{Carmalt2004} In the \ce{NbS2} system only the 2H-\ce{NbS2} polytype is known to be superconducting with a critical temperature of $T_{\rm c} \approx$ 6 K, although it is worth mentioning that there are reports that found also the 3R polytype to be superconducting with a very similar critical temperature.\cite{Naito1982,Onabe1978,Guillamon2008a, VanMaaren1966,Kacmarcik2010,Kacmarcik2010b} As we shall show later in this work, these observations of superconductivity in samples of the 3R-\ce{NbS2} polytype may likely be caused the presence of traces of 2H-\ce{NbS2}, which are challenging to identify by means of X-ray diffraction. The challenge to prepare phase pure samples of 2H-\ce{NbS2} and 3R-\ce{NbS2} has earlier been recognized by \textit{Fisher et al.}.\cite{Fisher1980}  There it was highlighted that the sulfur pressure during synthesis is crucial for the phase formation of the product.

Here, we investigate the reaction conditions for synthesizing the 2H-\ce{NbS2} and 3R-\ce{NbS2} polytypes under standardized, systematically altered parameters by means of solid-state synthesis. Our findings result in a detailed synthesis map for the whole \ce{NbS2} system. This synthesis map allows for identification of targeted synthesis conditions for the preparation of phase-pure samples in this system. Our analysis of the physical and superconducting properties reveals that specific-heat measurements are crucial for the identification of superconducting materials. This is especially true in this systems, since we can show that samples with mixed phases may easily be mistaken as bulk superconductors.

\section{Experimental Section}

All samples were prepared by means of high-temperature solid-state synthesis from the pure elements niobium (powder, Alfa Aesar, 99.99\%) and sulfur (pieces, Alfa Aesar, 99.999\%). The niobium and sulfur were mixed in their respective ratios, and thoroughly ground to a fine mixture and pressed into pellets. Each pellet was sealed in a quartz glass ampoule under 1/3 atm argon. The samples were then heated (with 180 \degree C/h) to temperatures ranging from $T$ = 600 \degree C to 950 \degree C for 3 days. 

Powder X-ray diffraction (PXRD) patterns were collected on an STOE STADI P diffractometer in transmission mode equipped with a Ge-monochromator using Cu K\textsubscript{$\alpha$1} radiation and on a Rigaku SmartLab in reflection mode using Cu K\textsubscript{$\alpha$} radiation. Scanning electron microscopy (SEM) was performed on a Zeiss Supra 50 VP. 

The temperature-dependent magnetization was measured in a Quantum Design magnetic properties measurement system (MPMS) equipped with a 7 Tesla (T) magnet and with a reciprocating sample option (RSO). The samples were measured in a gelatin capsule, where the layered materials naturally arranged perpendicular to the external magnetic field. The measurements were performed upon warming the sample in zero-field mode. The specific-heat capacity measurements were performed in a Quantum Design EverCool physical property measurement system (PPMS) equipped with a 9 T magnet. These measurements were performed with the Quantum Design heat-capacity option using a relaxation technique. 

SEM images were taken with a JEOL JSM 6060 scanning electron microscope and the elemental composition analysed using energy dispersive X-ray (EDX) (Bruker axes) attached to the JEOL JSM 6060. 

\section{Result and Discussion}
\subsection{Synthesis and X-ray Diffraction}
Crystallographic parameters of the 3R-\ce{NbS2} and 2H-\ce{NbS2} polytypes are given in table \ref{tbl:example}. 

\begin{figure}[H]
\centering
  \includegraphics[width=0.8\linewidth]{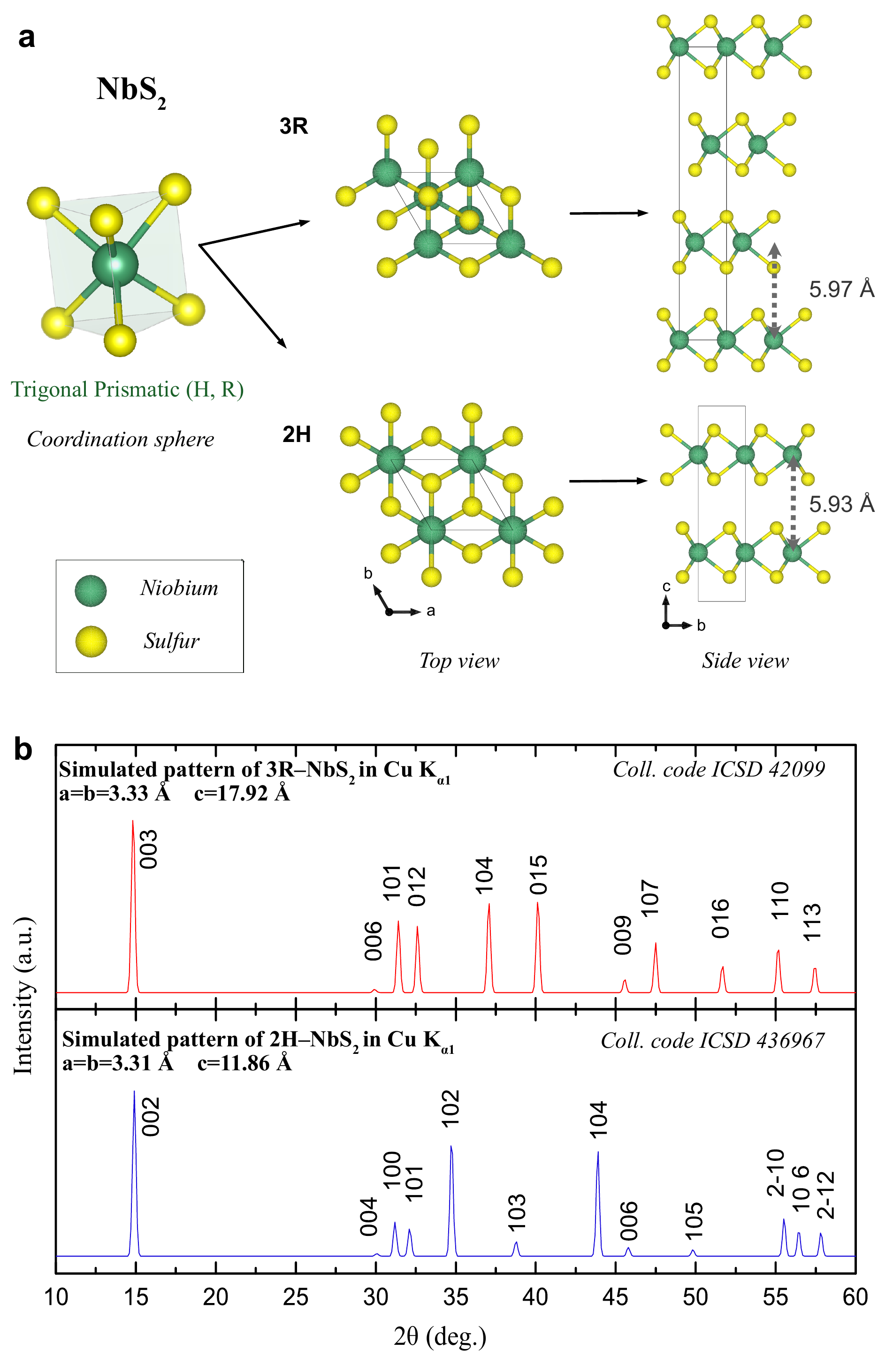}
  \caption{ (a) Polytypes of \ce{NbS2}. Left side: illustration of the trigonal prismatic coordination sphere, later resulting to the 3R-\ce{NbS2} and 2H-\ce{NbS2}. Middle: top view, and right: side view showing the number of van-der-Waals layers per unit cells. In green the metal Nb, in yellow the chalcogenide S. (b) Simulated PXRD patterns of 3R-\ce{NbS2} in red and 2H-\ce{NbS2} in blue with Cu \(K_{\alpha 1}\) radiation, including the respective Miller indices of the reflexions.}
  \label{Figure1}
\end{figure}

\begin{table}[h]
\small
  \caption{\ Summary of the crystallographic parameters for \ce{NbS2}.}
  \label{tbl:example}
  \begin{tabular*}{0.48\textwidth}{@{\extracolsep{\fill}} c c c}
    \hline
      & 3R-\ce{NbS2} & 2H-\ce{NbS2}   \\
    \hline
    Space group     & R3\textit{m} (No. 160) & $P6_{\rm 3}$/\textit{mmc} (No. 194) \\
    Z & 3 & 2 \\
    $\#$ of layers per unit cell & 3 & 2\\
    stacking sequence& ABC &AB \\
    \hline
  \end{tabular*}
\end{table}
Figure \ref{Figure1}(a) shows their geometry and crystal structure. 
The fundamental trigonal prismatic building block [\ce{NbS6}] is shown, which leads to the hexagonal (H) and rhombohedral (R) polytypes, when arranged in layers and stacked accordingly. Specifically, the views along the [001] and [100] direction of the crystal structure of the two stable 3R-\ce{NbS2} and 2H-\ce{NbS2} polytypes are shown with the highlighted unit cell in black. Whereas the covalent bonding within a layer is of trigonal primatic geometry for both materials, their stacking sequence differs, resulting in the different respective polytypes. The simulated PXRD patterns of the two compounds are shown in Figure \ref{Figure1}(b). These isotropic PXRD patterns have reflection positions that are very similar because of their common sublattice. Furthermore, the reflections that can be clearly distinct from each other are challenging to observe for real preferentially oriented, anisotropic samples, as we will discuss in detail below.

Polycrystalline samples of \ce{NbS2} were prepared under standardized conditions by means of conventional solid state synthesis. Each synthesis was performed (i) with a total mass of $m $= 400 mg of the reactants, (ii) in quartz glass ampoules (standing upright in the muffle furnace) of precise length of $l =$75 mm, a diameter of $d_{\rm wall} =$8 mm, and a wall thickness of $d_{\rm thickness} =$ 1 mm, (iii) with a heating rate from room-temperature to the final temperature of 180 $^\circ$C/h, for a total heating duration of precisely 72 h, (iv) all samples were eventually quenched into water after the reaction. Subsequent quenching of the samples in water proved to be important in order to remove excess of residual sulfur. Especially the samples synthesized with a sulfur excess had residual yellow unreacted sulfur at the top of the quartz tubes, well separated from the dark grey \ce{NbS2} products. 

A total of 56 samples were prepared and analysed by means of powder X-ray diffraction, resulting from different synthesis attempts of varying nominal stoichiometries in NbS$_{\rm x}$ with $x$ ranging from 1.7 to 2.3 in 0.1 steps and synthesis temperatures ranging between 600 \degree C to 950 \degree C in 50 \degree C steps. Three samples oxidized in the process, which is rendered by three missing points in figure \ref{Figure3}. No impurities of side-products or the starting materials were observed in any of these samples. All resulting products were dark grey. The highly crystalline samples had all a metallic luster, distinguishing them from their amorphous counterpart. An analysis of the morphology of the samples synthesized at 2.3 eq for various temperatures was done by means of SEM (see SI). At low synthesis temperatures, no distinct shape of crystals is formed, whereas we obtain platy crystals at higher synthesis temperatures.

\begin{figure}
\centering
  \includegraphics[width=0.9\linewidth]{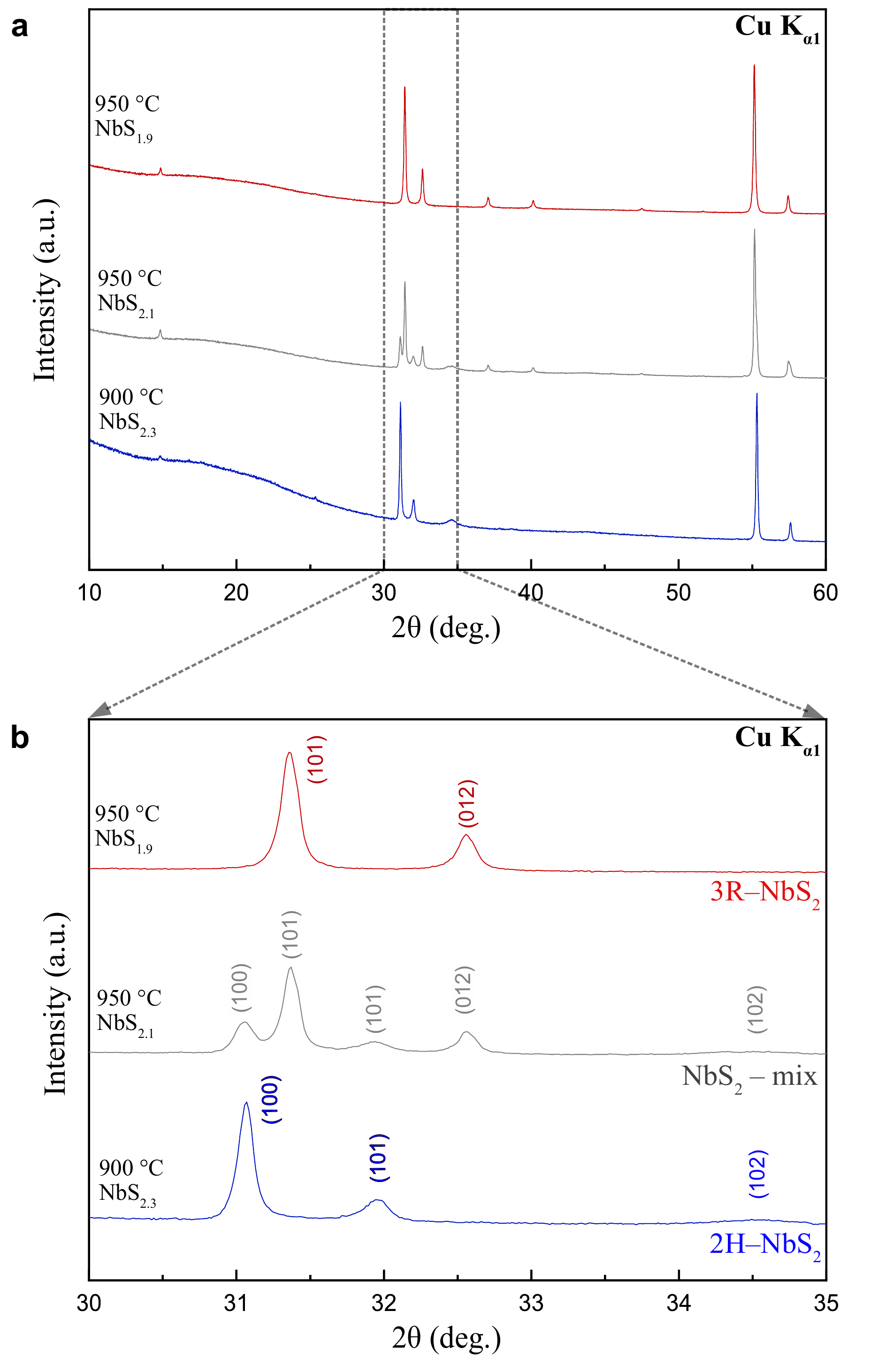}
  \caption{Powder x-ray diffraction patterns for different synthesis temperatures and compositions resulting in the different polytypes. (a) PXRD patterns obtained in the transmission mode using Cu \(K_{\alpha 1}\) radiation over a 2$\theta$ range of 10\degree \ to 60\degree. Clearly observable is the strong intensity of the hk0 reflections, due to the preferred orientation of the samples. (b) Zoom in the 2$\theta$ = 30 - 35\degree, showing the h0l and 0kl reflections, allowing the distinction of the polytypes. }
  \label{Figure2}
\end{figure}

Stoichiometries and/or synthesis temperatures outside of these specific conditions lead to the formation of considerable amounts of impurities. PXRD patterns for all samples are shown in the Supplemental Information. In order to illustrate the resulting differences between them, three representative patterns are shown in Figure \ref{Figure2}. These three pattern correspond to a phase-pure sample of 3R-\ce{NbS2} (red line), a phase pure sample of 2H-\ce{NbS2} (blue line), and a mixed sample, containing both polytypes (grey line). These three samples were also the ones that later were used for physical properties measurements (see below). It should be noted that the PXRD pattern look, at first glance, remarkably similar, hence a detailed analysis is required to accentuate the differences.

In PXRD, preferred orientation creates a systematic error in the observed intensities of diffraction peaks. The platy shape of the two \ce{NbS2} polytypes poses a challenge for obtaining an unbiased method to distinguish them by means of X-ray diffraction.  In the Bragg-Brentano geometry, i.e. reflection mode, the intensity of the 00l reflections will be heavily increased, because the (001) planes are oriented in such a way to be in reflection condition with the diffractometer. In the Debye-Scherrer geometry, i.e. transmission mode, the X-rays pass through the platelets and thus the intensities of the hk0 reflections will be heavily increased, since the (hk0) planes are perpendicular to the (001) planes. A comparison of the obtained PXRD patterns for the same sample on the two different instrument modes for the 2H-\ce{NbS2} polytype is given in the Supplemental Information. 

Both polytypes are crystallizing in a hexagonal setting. Equation \ref{hexagonal} helps to calculate the $d_{\rm hkl}$.
\begin{equation}
\dfrac{1}{d_{\rm hkl}}=\dfrac{4}{3} (\dfrac{h^2 + k^2 +hk}{a^2}) + \frac{l^2}{c^2}
\label{hexagonal}
\end{equation}
It is true that their c axes follow the relation $c_{\rm 2H} = \dfrac{2}{3} c_{\rm 3R}$ while showing similar a parameter. Therefore, the position of the 00l reflections originating from the crystal planes parallel to the layers will remain invariant: $d_{\rm 002n}(2H)=d_{\rm 003n}(3R)$ with n being an integer. The rhombohedral centering of the 3R polytype in hexagonal setting will show only every third reflection on the [001] axis (003, 006, 009) and the 2H polytype, because of its  6$_{\rm (3)}$ screw rotation, every second reflection on this axis (002, 004, 006), both at identical 2$\theta$ positions.\\
The hk0 reflections originating of the crystal planes perpendicular to the layers of both \ce{NbS2} polytypes will also remain invariant, because of the same length of a. Thus both polytypes can thus be only distinguished from one another by the position and intensities of h0l, 0kl and hkl reflections, which is particularly difficult since the intensities of these reflections are least pronounced. They are still observable in transmission mode, hence, here all samples were analysed by means of PXRD of the Debye-Scherrer geometry. 

In Figure \ref{Figure2}(b), we show a zoom-in for the a 2$\theta$ range of 30-35 \degree, where these reflections are most pronounced. 
 The 100 and 012 reflections of 3R-\ce{NbS2}, and the 100 and 101 of 2H-\ce{NbS2} allow for the differentiation of the two polytypes. Especially in the sample containing a mixture of both the presence of all four of these reflections - arising from both of the two polytypes - becomes most apparent in direct comparison. We can state that there is a substantial amount of the 3R-\ce{NbS2} polytype present in this sample, however a quantitative analysis of the amounts of the different phases is not possible with a PXRD analysis, due to the preferred orientation. Simply comparing the 100 of the 2H polytype with the 101 reflections of the 3R-\ce{NbS2} polytype at 31\degree and 31.5\degree might give the wrong impression that the 2H-\ce{NbS2} is the minority phase, while it actually is the majority (approximately 75~\%) phase, as we will argue below. It should be noted that the 102 reflection of 2H-\ce{NbS2} in any of the obtained samples is comparably broad, which may likely be originating from stacking faults or turbostratic disorder arising from the random orientation of successive layers about the stacking direction.

\subsection{Synthesis Map of the \ce{NbS2} System}

\begin{figure*}
\centering
  \includegraphics[width = 0.7\textwidth]{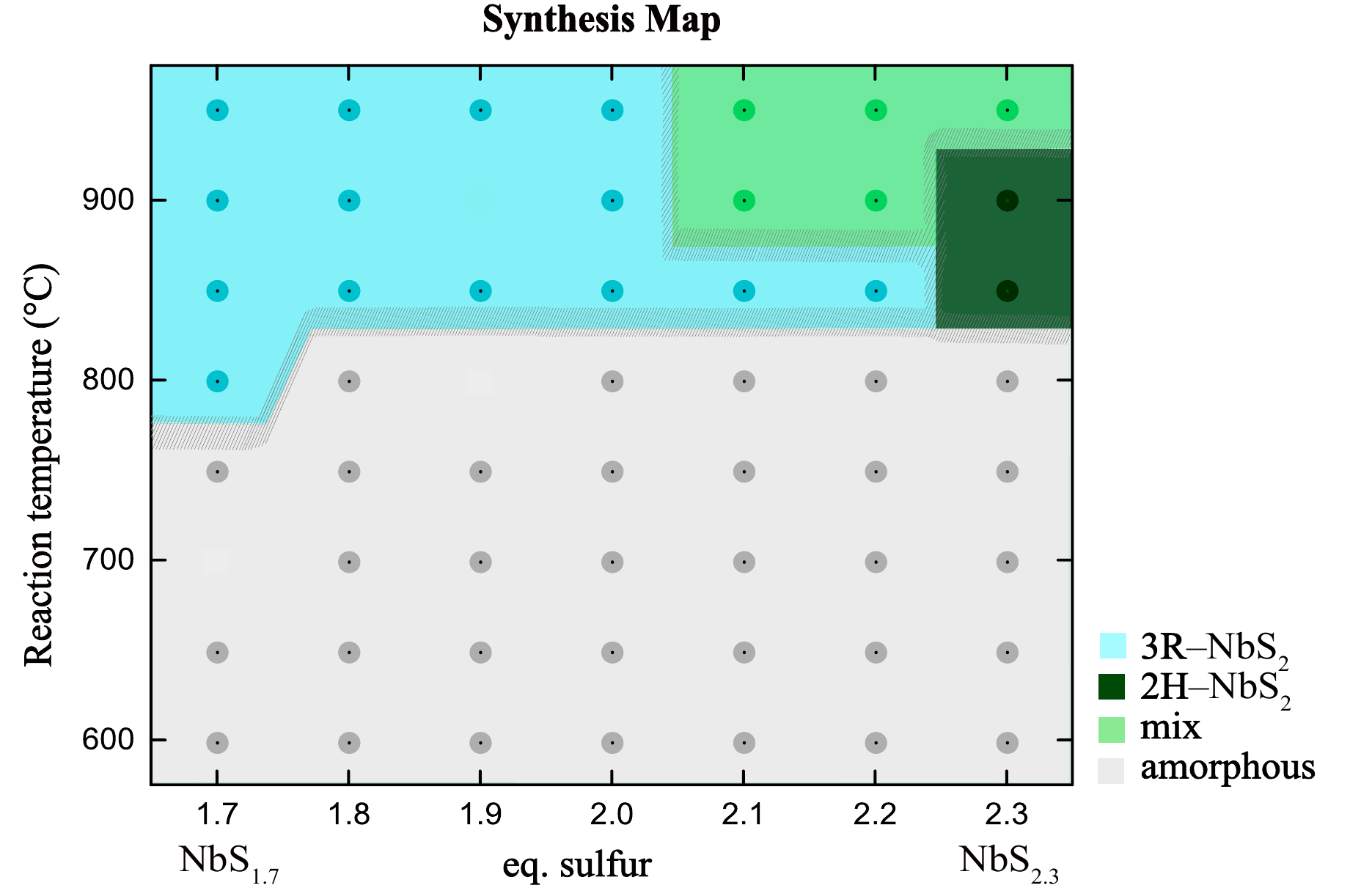}
  \caption{Synthesis map of the \ce{NbS2} system for \ce{NbS_x} with $x$ ranging from 1.7 to 2.3 in 0.1 steps and a synthesis temperature range between 600 \degree C to 950 \degree C in 50 \degree C steps. Each data point represents a synthesis according to the respective synthesis protocol described in the text. Phase pure samples of 3R-\ce{NbS2} in blue, phase-pure samples of 2H-\ce{NbS2} in dark green, the mixed-phase region is light green, and the amorphous region is marked in grey. The 2H-\ce{NbS2} polytype appears only in a in a very narrow synthesis window.}
  \label{Figure3}
\end{figure*}

The performed systematic syntheses and their respective analyses by means of PXRD in the \ce{NbS2} system, allow for the compilation of a comprehensive synthesis map, which is shown in Figure \ref{Figure3}. The careful analysis of the obtained PXRD pattern allowed us to distinguish between 4 different regions: (i) amorphous, (ii) phase pure 2H-\ce{NbS2}, (iii) phase pure 3R-\ce{NbS2}, and (iv) mixed phase regions of both the 2H and the 3R polytypes. The amorphous region \footnote{Here, "amorphous" is used as a collective term describing the non-crystalline and low-crystalline region of the synthesis map.} was here defined for samples with the FWHM being larger than 2$\theta$=0.21\degree for the reflection at 2$\theta \approx$ 14.8\degree, which corresponds to either the 002 reflection of 2H-\ce{NbS2}, or the 003 reflection of 3R-\ce{NbS2} polytype. For samples with larger values of FWHM the 2H-\ce{NbS2} and 3R-\ce{NbS2} polytypes could not be distinguished in an unbiased fashion. The gradual change from amorphous to crystalline samples can be especially well observed in the PXRD patterns of the samples at a fixed sulfur content (see Supplemental Information). An analysis of the morphology of the samples synthesized at 2.3 eq for various temperatures was done by means of SEM (see SI). The points in the synthesis map correspond to each synthesized sample and indicate the exact temperature of synthesis and sulfur content used. 
The formation of phase-pure 2H-\ce{NbS2} was only observed in a very narrow temperature and stoichiometry interval. Overall, we find that the 3R-\ce{NbS2} polytype preferentially forms in a stoichiometric or sulfur deficiency environment, whereas excess of it is needed for the formation of the 2H-\ce{NbS2} polytype. This observation is also in agreement with earlier findings by \textit{Fisher et al}, where substantial sulfur pressures were found to be crucial for stabilization of the 2H-\ce{NbS2} polytype. It might be speculated that the 3R polytype preferentially forms with a sulfur deficiency, because its ABC stacking reduces the likelihood of having two sulfur vacancies directly above or below each other. However, our systematic EDX analysis, for both the 2H-\ce{NbS2} and 3R-\ce{NbS2} polytypes, reveals samples very similar sulfur contents of 1.89 and 1.92, respectively (see SI). Our findings do not only affect the formation of polycrystalline samples, but they also have implications for the preparation of single crystals of the different \ce{NbS2} polytypes, as it is very likely that the two different phases can be intergrowth of each other, which is also called allotwins.

\subsection{Superconducting Properties of \ce{NbS2}}
\begin{figure}
\centering
  \includegraphics[width=0.8\linewidth]{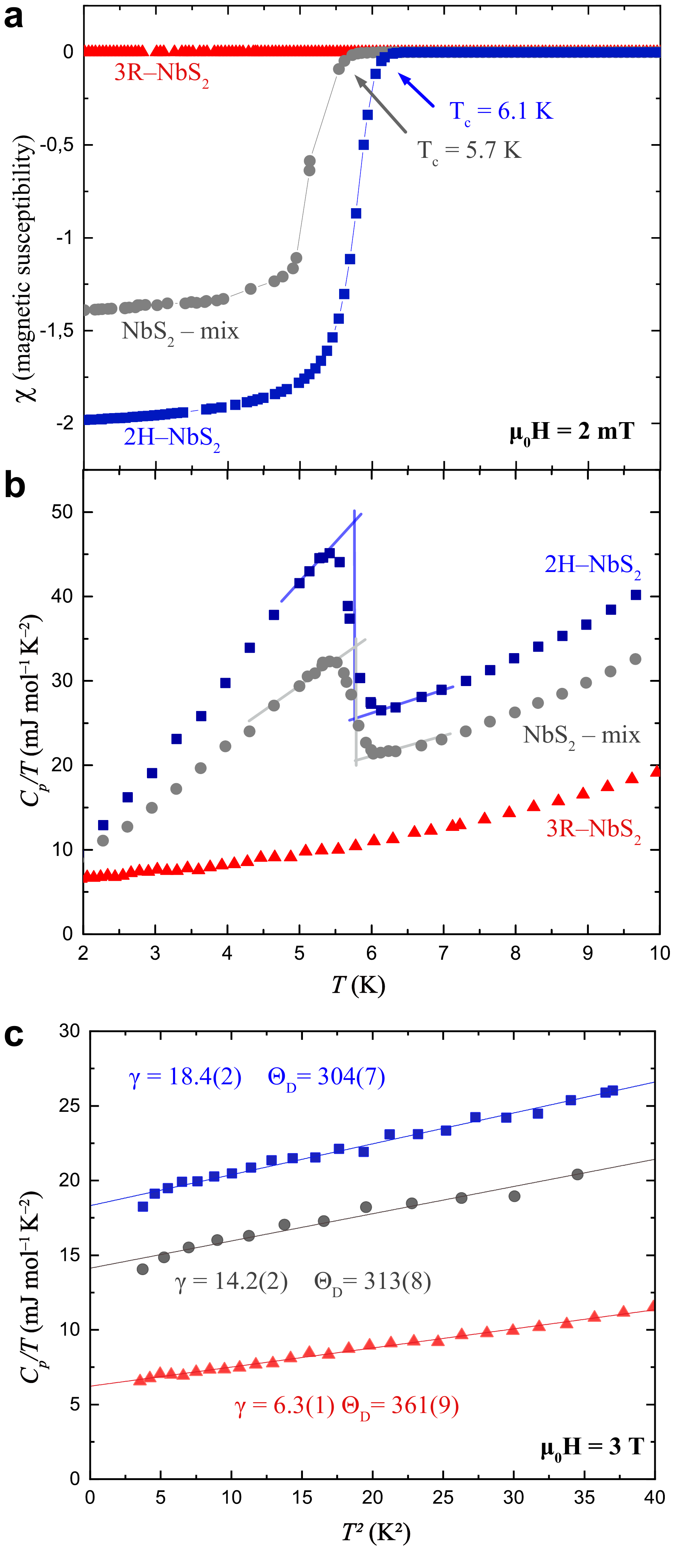}
  \caption{Physical and superconducting properties of a phase pure 2H-\ce{NbS2} (blue), a phase pure 3R-\ce{NbS2} (red), and a sample consisting of both polytypes (grey). (a) ZFC temperature-dependent magnetization $M$($T$) measured in a field of $\mu_0 H =$ 2 mT, in the vicinity of the superconducting transition. (b) Temperature-dependent specific heat $C$($T$)/$T$ in the vicinity of the superconducting transition, measured in zero applied field. The solid lines outline the entropy conserving construction. (c) $C$($T$)/$T$ in the normal state versus $T^2$. The solid lines are fits to Equation \ref{specificheat}. }
  \label{Figure4}
\end{figure}

Temperature-dependent magnetization and temperature-dependent specific-heat measurements were performed (see Figure \ref{Figure4}) to determine the superconducting properties of the obtained samples. Specifically, phase pure 2H-\ce{NbS2}, phase-pure 3R-\ce{NbS2}, and a sample with both polytypes were investigated (same sample as discussed in Figure \ref{Figure2}). The superconducting critical temperatures are found to be $T_{\rm c} =$ 6.1 K in the magnetization and 5.7 K in the specific heat for the phase-pure 2H-\ce{NbS2} polytype, and 5.7 K in the magnetization and the specific heat for the mixed phase sample. In neither measurement, we find any superconducting transition in the phase-pure 3R polytype sample above a temperature of $T >$ 1.75 K. Hence, previous reports of superconductivity in 3R-\ce{NbS2} were likely originating from 2H-\ce{NbS2} impurities, as we will show in the following.\cite{VanMaaren1966} 

In Figure \ref{Figure4}(a), we show the magnetization measurements as the unitless magnetic susceptibility $\chi = M/H$ for all three samples. The measurements were performed in a temperature range between $T =$~1.75 and 10~K in an external magnetic field of $\mu_0 H$~=~2 mT in zero-field cooled (ZFC) mode. A bulk superconductor is an ideal diamagnet in the Meissner state, hence, a magnetic susceptibility in the ZFC mode of $\chi =$ -1, corresponding to a 100 \% shielding fraction, is expected. At temperatures below $T$ = 2 K, the diamagnetic signal of 2H-\ce{NbS2} saturates at a value of nearly 200 \% of the shielding fraction. This value exceeds the theoretical value for an ideal diamagnet by a factor of approximately 2. This large shielding fraction is due to demagnetization effects. Thereby, the effective magnetic field is reduced due to a demagnetizing magnetic field $H_{\rm D}$, which in turn is generated by the magnetization $M$ within the superconductor according to

\begin{equation}
 H_{\rm in} = H_{\rm ext} - H_{\rm D} = H_{\rm ext} - {\rm n}M = \left( \dfrac{1}{1-n} \right)H_{\rm ext}
\end{equation}

with n being the so-called demagnetizing factor. For geometrically, well-defined cases such as e.g. ellipsoids or plates $H_{\rm D}$ is linearly related to the magnetization $M$ by a constant. The demagnetization fields are commonly more challenging to calculate, especially for arbitrarily shaped real objects. For the extreme case of an ideal diamagnet in the shape of a plate, which is placed perpendicular to an external magnetic field $H_{\rm ext}$ the demagnetizing factor approaches unity. Here, for the measurements of polycrystalline samples of layered TMDs, the plate-like samples naturally arranged perpendicular to the external magnetic field, which enhances the observed diamagnetic shielding fraction heavily. This effect also enhances the shielding fraction of the sample containing a mixture of both polytypes 2H-\ce{NbS2} and 3R-\ce{NbS2}, leading to a measured shielding fraction of nearly 140 \% at $T$ = 1.75 K. This sample might be easily mistaken for a bulk superconducting sample, due to this large volume fraction, the well-defined, sharp superconducting transition, and the critical temperature of $T_{\rm c}$ = 5.7 K, which is in the range of values reported in the literature for superconductors in the \ce{NbS2} system.

Therefore, these superconductors with a layered crystal structure resulting in a platy crystal shape are a particularly characteristic example of how magnetic susceptibility measurements, and in extension also resistivity measurements -- which are showing a state of zero-resistance, even at very low concentrations of superconducting grains -- are insufficient for the characterization and confirmation of bulk superconductors. Rather true bulk measurements are needed, especially the measurement of the temperature-dependent specific heat $C$($T$). At $T = T_{\rm c}$ the specific heat of the paired electrons is larger than the specific heat of the electrons in the normal state 

\begin{equation}
C_{\rm super} (T_{\rm c}) \quad > \quad C_{\rm el} (T_{\rm c}).
\end{equation}

This leads to a characteristic discontinuity at the critical temperature, which according to the Bardeen-Cooper-Schrieffer (BCS) theory\cite{Bardeen1957} is 

\begin{equation}
    \left( \dfrac {C_{\rm super}-C_{\rm el}}{\gamma {T_{\rm c}}}\right)_{T_{\rm c}} = 1.43.
\end{equation}

Values close or larger than these 1.43 for the discontinuity in the specific heat are a strong indicator, and believed to be proof for bulk superconductivity. This difference means that less electrons are forming Cooper-pairs, then generally would be expected from the density of electronic states at the Fermi level $D$($E_{\rm Fermi}$).

In Figure \ref{Figure4}(b), we show the temperature-dependent specific heat $C$($T$)/$T$ in a temperature range between $T$ = 2 and 10 K for phase pure 2H-\ce{NbS2}, phase pure 3R-\ce{NbS2}, and the sample consisting of both polytypes. A well-pronounced, sharp discontinuity at the transition to the superconducting state is observed for the 2H-\ce{NbS2} polytype, as well as for the mixed sample. 

In Figure \ref{Figure4}(c), we show the analysis of the normal state of the three samples. The normal state specific heat contributions have been fitted to the data according to the general expression

\begin{equation}
\label{specificheat}
\frac{C(T)}{T} = \gamma + \beta T^2
\end{equation}

with the Sommerfeld constant $\gamma$ and $\beta$ = 12$\pi^4$nR/5$\Theta_D^3$, where n is the number of atoms per formula unit, R is the gas constant, and $\Theta_D$ is the Debye temperature. 

We find for the phase pure sample of 2H-\ce{NbS2} an approximately 3 times larger Sommerfeld constant of $\gamma_{\rm 2H} =$ 18.4(2) mJ mol$^{-1}$ K$^{-2}$ than the one of 3R-\ce{NbS2} with $\gamma_{\rm 3R} =$ 6.3(1) mJ mol$^{-1}$ K$^{-2}$. This may likely explain, why the 2H-\ce{NbS2} polytype is a superconductor, while the 3R-\ce{NbS2} polytype is not. This large difference corresponds to a much higher density of states at the Fermi level $D$($E_{\rm Fermi}$) for the 2H-\ce{NbS2} polytype. According to the BCS theory the critical temperature is proportional to density of states at the Fermi level $D(E_{\rm Fermi}) \propto T_{\rm c}$, which in turn explains the absence of superconductivity in 3R-\ce{NbS2} above $T =$ 1.75 K. It is noteworthy that also the phononic contribution between the two polytypes differ, resulting in different Debye temperatures $\Theta_{\rm D}$ of 304(7) K and 361(9) K for the 2H-\ce{NbS2} and 3R-\ce{NbS2} polytypes, respectively. This is surprising, as the basic building block, i.e. the monolayer of \ce{NbS2} is for both polytypes the same, hence the electronic and phononic differences must be caused by the different stackings of the layers, i.e. the electronic and phononic overlap through the van-der-Waals gaps. A possible explanation is the enhanced orbital overlap in the 2H-\ce{NbS2} due to its AB stacking, allowing for a higher fraction of atoms to be directly above or below other atoms.

The entropy-conserving constructions of the superconducting specific heat discontinuity are shown for the phase pure 2H-\ce{NbS2} sample and the mixed-phase sample in Figure \ref{Figure4}(b) resulting in values for $\Delta C/\gamma T_{\rm c}$ of 1.30 and 0.98, for the phase-pure 2H-\ce{NbS2} and the mixed sample, respectively. The value $\Delta C/\gamma T_{\rm c}$ for phase-pure 2H-\ce{NbS2} is in excellent agreement with earlier studies on high-quality single crystals of 2H-\ce{NbS2}.\cite{Kacmarcik2010,Kacmarcik2010b} Since the specific heat is truly a bulk measure, we can state that the mixed sample with $\Delta C$ almost being parity to $\gamma T_{\rm c} $, contains of maximally \ 75 \% of the 2H-\ce{NbS2} polytype, but it might nevertheless be easily mistaken for a bulk superconductor. This finding is in agreement with earlier reports for other superconducting systems, where specific heat measurements were also found to be crucial for the identification of bulk superconducting phases \cite{Carnicom2019,VonRohr2014}. A summary of the obtained superconducting parameters of both samples is given in table \ref{tbl:example1}.

\begin{table}[h]
\small
  \caption{\ Summary of the physical and superconducting properties of \ce{NbS2}.}
  \label{tbl:example1}
  \begin{tabular*}{0.48\textwidth}{@{\extracolsep{\fill}} c c c c c c}
    \hline
    
    sample  & $\gamma$  & $\Theta_{\rm D} $ & $T_{\rm c,heat}$  & $T_{\rm c,mag} $ & $\Delta C/\gamma T_{\rm c}$ \\
            &(mJ mol$^{-1}$ K$^{-2}$) &(K)& (K) &(K) &\\
    \hline
    
    3R--\ce{NbS2} & 6.3(1) & 361(9) & -- & -- & --\\
    2H--\ce{NbS2} & 18.4(2) & 304(7) & 5.7 & 6.1 & 1.30\\
    mixed sample & 14.2(2) & 313(8) & 5.7 & 5.7 & 0.98\\
    \hline
  \end{tabular*}
\end{table}

\section{Summary and Conclusions}
In summary, we have performed a series standardized solid-state syntheses in the \ce{NbS2} system, which allowed us to establish a comprehensive synthesis map for the formation of the two polytypes 2H-\ce{NbS2} and 3R-\ce{NbS2}. We show that the distinguishing of the two polytypes by PXRD is not trivial, as the differing reflections are least pronounced due to preferred orientation leading to systematic errors. Furthermore, we find that there is a relatively large stoichiometry and synthesis-temperature region, where mixed samples consisting of both polytypes are formed and that there are well-defined synthesis conditions that lead to phase pure samples of either polytype.

Our physical property measurements on a phase-pure sample of 3R-\ce{NbS2}, on a phase-pure sample of 2H-\ce{NbS2}, and a mixed phase sample have confirmed earlier reports that 2H-\ce{NbS2} is a bulk superconductor. We show that 3R-\ce{NbS2} is not a superconductor above $T =$ 1.75 K, in contrary to some earlier reports. We, furthermore, find that from magnetization measurements the mixed sample may easily be mistaken for a bulk superconducting sample. Hence, specific heat measurements, as true bulk measurements, are found to be crucial for the identification of superconducting materials in this and related systems. For phase-pure 2H-\ce{NbS2}, we report a value of $\Delta C/\gamma T_{\rm c} =$ 1.30. Using specific heat measurements, and especially this value -- rather than magnetization and resistivity measurements -- as a measure for the purity of 2H-\ce{NbS2} single crystals, is found to be critical, as also single crystals might show substantial 3R-\ce{NbS2} inclusions (see, e.g. references \citenum{Leroux2012b,Katzke2002}). These 3R-\ce{NbS2} polytype inclusions may even occur in apparently large single crystals of 2H-\ce{NbS2}, due to the similar chemistry of the two polytypes. Therefore, inter-growth regions of the two polytypes may be mistaken for stacking faults. 

We conclude by pointing out that for the investigation of van-der-Waals materials in the \ce{NbS2} system -- but also in chemically related systems -- great care has to be taken on choosing the right synthesis conditions for obtaining phase pure samples, since the formation of impurity phases is likely, and can even more likely be overlooked due to structural similarities that cause similar PXRD patterns. 

\section*{Conflicts of interest}
There are no conflicts to declare.

\section*{Acknowledgements}
The authors thank Manuele Balestra and Mark Blumer for help during the synthesis. The authors thank Dr. Robin Lefèvre for helpful discussions. This  work  was  supported  by  the  Swiss National Science Foundation under Grant No. PZ00P2\_174015. Work at GUT was supported by the National Science Centre (Poland), grant number: UMO-2017/27/B/ST5/03044. This research is also funded by the Swedish Research Council (VR) through a neutron project grant Dnr. 2016-06955.

\balance

\bibliography{nbs2paper.bib} 

\providecommand*{\mcitethebibliography}{\thebibliography}
\csname @ifundefined\endcsname{endmcitethebibliography}
{\let\endmcitethebibliography\endthebibliography}{}
\begin{mcitethebibliography}{33}
\providecommand*{\natexlab}[1]{#1}
\providecommand*{\mciteSetBstSublistMode}[1]{}
\providecommand*{\mciteSetBstMaxWidthForm}[2]{}
\providecommand*{\mciteBstWouldAddEndPuncttrue}
  {\def\EndOfBibitem{\unskip.}}
\providecommand*{\mciteBstWouldAddEndPunctfalse}
  {\let\EndOfBibitem\relax}
\providecommand*{\mciteSetBstMidEndSepPunct}[3]{}
\providecommand*{\mciteSetBstSublistLabelBeginEnd}[3]{}
\providecommand*{\EndOfBibitem}{}
\mciteSetBstSublistMode{f}
\mciteSetBstMaxWidthForm{subitem}
{(\emph{\alph{mcitesubitemcount}})}
\mciteSetBstSublistLabelBeginEnd{\mcitemaxwidthsubitemform\space}
{\relax}{\relax}

\bibitem[Coleman \emph{et~al.}(2011)Coleman, Lotya, O'Neill, Bergin, King,
  Khan, Young, Gaucher, De, Smith, Shvets, Arora, Stanton, Kim, Lee, Kim,
  Duesberg, Hallam, Boland, Wang, Donegan, Grunlan, Moriarty, Shmeliov,
  Nicholls, Perkins, Grieveson, Theuwissen, McComb, Nellist, and
  Nicolosi]{Coleman2011a}
J.~N. Coleman, M.~Lotya, A.~O'Neill, S.~D. Bergin, P.~J. King, U.~Khan,
  K.~Young, A.~Gaucher, S.~De, R.~J. Smith, I.~V. Shvets, S.~K. Arora,
  G.~Stanton, H.-Y. Kim, K.~Lee, G.~T. Kim, G.~S. Duesberg, T.~Hallam, J.~J.
  Boland, J.~J. Wang, J.~F. Donegan, J.~C. Grunlan, G.~Moriarty, A.~Shmeliov,
  R.~J. Nicholls, J.~M. Perkins, E.~M. Grieveson, K.~Theuwissen, D.~W. McComb,
  P.~D. Nellist and V.~Nicolosi, \emph{Science (80-. ).}, 2011, \textbf{331},
  568--571\relax
\mciteBstWouldAddEndPuncttrue
\mciteSetBstMidEndSepPunct{\mcitedefaultmidpunct}
{\mcitedefaultendpunct}{\mcitedefaultseppunct}\relax
\EndOfBibitem
\bibitem[Lotsch(2015)]{Lotsch2015b}
B.~V. Lotsch, \emph{Annu. Rev. Mater. Res.}, 2015, \textbf{45}, 85--109\relax
\mciteBstWouldAddEndPuncttrue
\mciteSetBstMidEndSepPunct{\mcitedefaultmidpunct}
{\mcitedefaultendpunct}{\mcitedefaultseppunct}\relax
\EndOfBibitem
\bibitem[Ryu \emph{et~al.}(2018)Ryu, Chen, Kim, Tsai, Tang, Jiang, Liou, Kahn,
  Jia, Omrani,\emph{et~al.}]{Ryu2018}
H.~Ryu, Y.~Chen, H.~Kim, H.-Z. Tsai, S.~Tang, J.~Jiang, F.~Liou, S.~Kahn,
  C.~Jia, A.~A. Omrani \emph{et~al.}, \emph{Nano letters}, 2018, \textbf{18},
  689--694\relax
\mciteBstWouldAddEndPuncttrue
\mciteSetBstMidEndSepPunct{\mcitedefaultmidpunct}
{\mcitedefaultendpunct}{\mcitedefaultseppunct}\relax
\EndOfBibitem
\bibitem[Xi \emph{et~al.}(2015)Xi, Zhao, Wang, Berger, Forr{\'{o}}, Shan, and
  Mak]{Xi2015a}
X.~Xi, L.~Zhao, Z.~Wang, H.~Berger, L.~Forr{\'{o}}, J.~Shan and K.~F. Mak,
  \emph{Nat. Nanotechnol.}, 2015, \textbf{10}, 765--769\relax
\mciteBstWouldAddEndPuncttrue
\mciteSetBstMidEndSepPunct{\mcitedefaultmidpunct}
{\mcitedefaultendpunct}{\mcitedefaultseppunct}\relax
\EndOfBibitem
\bibitem[Devarakonda \emph{et~al.}(2019)Devarakonda, Inoue, Fang,
  Ozsoy-Keskinbora, Suzuki, Kriener, Fu, Kaxiras, Bell, and
  Checkelsky]{Devarakonda2019}
A.~Devarakonda, H.~Inoue, S.~Fang, C.~Ozsoy-Keskinbora, T.~Suzuki, M.~Kriener,
  L.~Fu, E.~Kaxiras, D.~C. Bell and J.~G. Checkelsky, \emph{arXiv preprint
  arXiv:1906.02065}, 2019\relax
\mciteBstWouldAddEndPuncttrue
\mciteSetBstMidEndSepPunct{\mcitedefaultmidpunct}
{\mcitedefaultendpunct}{\mcitedefaultseppunct}\relax
\EndOfBibitem
\bibitem[Wang \emph{et~al.}(2017)Wang, Huang, Lin, Cui, Chen, Zhu, Liu, Zeng,
  Zhou, Yu,\emph{et~al.}]{Wang2017}
H.~Wang, X.~Huang, J.~Lin, J.~Cui, Y.~Chen, C.~Zhu, F.~Liu, Q.~Zeng, J.~Zhou,
  P.~Yu \emph{et~al.}, \emph{Nature communications}, 2017, \textbf{8},
  1--8\relax
\mciteBstWouldAddEndPuncttrue
\mciteSetBstMidEndSepPunct{\mcitedefaultmidpunct}
{\mcitedefaultendpunct}{\mcitedefaultseppunct}\relax
\EndOfBibitem
\bibitem[Xing \emph{et~al.}(2017)Xing, Zhao, Shan, Zheng, Zhang, Fu, Liu, Tian,
  Xi, Liu,\emph{et~al.}]{Xing2017}
Y.~Xing, K.~Zhao, P.~Shan, F.~Zheng, Y.~Zhang, H.~Fu, Y.~Liu, M.~Tian, C.~Xi,
  H.~Liu \emph{et~al.}, \emph{Nano Letters}, 2017, \textbf{17},
  6802--6807\relax
\mciteBstWouldAddEndPuncttrue
\mciteSetBstMidEndSepPunct{\mcitedefaultmidpunct}
{\mcitedefaultendpunct}{\mcitedefaultseppunct}\relax
\EndOfBibitem
\bibitem[von Rohr \emph{et~al.}(2019)von Rohr, Orain, Khasanov, Witteveen,
  Shermadini, Nikitin, Chang, Wieteska, Pasupathy, Hasan, Amato, Luetkens,
  Uemura, and Guguchia]{VonRohr2019a}
F.~O. von Rohr, J.-C. Orain, R.~Khasanov, C.~Witteveen, Z.~Shermadini,
  A.~Nikitin, J.~Chang, A.~R. Wieteska, A.~N. Pasupathy, M.~Z. Hasan, A.~Amato,
  H.~Luetkens, Y.~J. Uemura and Z.~Guguchia, \emph{Sci. Adv.}, 2019,
  \textbf{5}, eaav8465\relax
\mciteBstWouldAddEndPuncttrue
\mciteSetBstMidEndSepPunct{\mcitedefaultmidpunct}
{\mcitedefaultendpunct}{\mcitedefaultseppunct}\relax
\EndOfBibitem
\bibitem[Ribak \emph{et~al.}(2020)Ribak, Skiff, Mograbi, Rout, Fischer, Ruhman,
  Chashka, Dagan, and Kanigel]{Ribak2020}
A.~Ribak, R.~M. Skiff, M.~Mograbi, P.~Rout, M.~Fischer, J.~Ruhman, K.~Chashka,
  Y.~Dagan and A.~Kanigel, \emph{Science advances}, 2020, \textbf{6},
  eaax9480\relax
\mciteBstWouldAddEndPuncttrue
\mciteSetBstMidEndSepPunct{\mcitedefaultmidpunct}
{\mcitedefaultendpunct}{\mcitedefaultseppunct}\relax
\EndOfBibitem
\bibitem[Brown and Beerntsen(1965)]{Brown1965}
B.~E. Brown and D.~J. Beerntsen, \emph{Acta Crystallographica}, 1965,
  \textbf{18}, 31--36\relax
\mciteBstWouldAddEndPuncttrue
\mciteSetBstMidEndSepPunct{\mcitedefaultmidpunct}
{\mcitedefaultendpunct}{\mcitedefaultseppunct}\relax
\EndOfBibitem
\bibitem[Luo \emph{et~al.}(2015)Luo, Xie, Tao, Inoue, Gyenis, Krizan, Yazdani,
  Zhu, and Cava]{Luo2015}
H.~Luo, W.~Xie, J.~Tao, H.~Inoue, A.~Gyenis, J.~W. Krizan, A.~Yazdani, Y.~Zhu
  and R.~J. Cava, \emph{Proceedings of the National Academy of Sciences}, 2015,
  \textbf{112}, E1174--E1180\relax
\mciteBstWouldAddEndPuncttrue
\mciteSetBstMidEndSepPunct{\mcitedefaultmidpunct}
{\mcitedefaultendpunct}{\mcitedefaultseppunct}\relax
\EndOfBibitem
\bibitem[Qi \emph{et~al.}(2016)Qi, Naumov, Ali, Rajamathi, Schnelle, Barkalov,
  Hanfland, Wu, Shekhar, Sun, S{\"{u}}{\ss}, Schmidt, Schwarz, Pippel, Werner,
  Hillebrand, F{\"{o}}rster, Kampert, Parkin, Cava, Felser, Yan, and
  Medvedev]{Qi2016a}
Y.~Qi, P.~G. Naumov, M.~N. Ali, C.~R. Rajamathi, W.~Schnelle, O.~Barkalov,
  M.~Hanfland, S.~C. Wu, C.~Shekhar, Y.~Sun, V.~S{\"{u}}{\ss}, M.~Schmidt,
  U.~Schwarz, E.~Pippel, P.~Werner, R.~Hillebrand, T.~F{\"{o}}rster,
  E.~Kampert, S.~Parkin, R.~J. Cava, C.~Felser, B.~Yan and S.~A. Medvedev,
  \emph{Nat. Commun.}, 2016, \textbf{7}, 1--7\relax
\mciteBstWouldAddEndPuncttrue
\mciteSetBstMidEndSepPunct{\mcitedefaultmidpunct}
{\mcitedefaultendpunct}{\mcitedefaultseppunct}\relax
\EndOfBibitem
\bibitem[Keum \emph{et~al.}(2015)Keum, Cho, Kim, Choe, Sung, Kan, Kang, Hwang,
  Kim, Yang, Chang, and Lee]{Keum2015}
D.~H. Keum, S.~Cho, J.~H. Kim, D.-H. Choe, H.-J. Sung, M.~Kan, H.~Kang, J.-Y.
  Hwang, S.~W. Kim, H.~Yang, K.~J. Chang and Y.~H. Lee, \emph{Nat. Phys.},
  2015, \textbf{11}, 482--486\relax
\mciteBstWouldAddEndPuncttrue
\mciteSetBstMidEndSepPunct{\mcitedefaultmidpunct}
{\mcitedefaultendpunct}{\mcitedefaultseppunct}\relax
\EndOfBibitem
\bibitem[Guguchia \emph{et~al.}(2017)Guguchia, Von~Rohr, Shermadini, Lee,
  Banerjee, Wieteska, Marianetti, Frandsen, Luetkens,
  Gong,\emph{et~al.}]{Guguchia2017}
Z.~Guguchia, F.~Von~Rohr, Z.~Shermadini, A.~Lee, S.~Banerjee, A.~Wieteska,
  C.~Marianetti, B.~Frandsen, H.~Luetkens, Z.~Gong \emph{et~al.}, \emph{Nature
  communications}, 2017, \textbf{8}, 1--8\relax
\mciteBstWouldAddEndPuncttrue
\mciteSetBstMidEndSepPunct{\mcitedefaultmidpunct}
{\mcitedefaultendpunct}{\mcitedefaultseppunct}\relax
\EndOfBibitem
\bibitem[Guguchia \emph{et~al.}(2018)Guguchia, Kerelsky, Edelberg, Banerjee,
  von Rohr, Scullion, Augustin, Scully, Rhodes,
  Shermadini,\emph{et~al.}]{Guguchia2018}
Z.~Guguchia, A.~Kerelsky, D.~Edelberg, S.~Banerjee, F.~von Rohr, D.~Scullion,
  M.~Augustin, M.~Scully, D.~A. Rhodes, Z.~Shermadini \emph{et~al.},
  \emph{Science advances}, 2018, \textbf{4}, eaat3672\relax
\mciteBstWouldAddEndPuncttrue
\mciteSetBstMidEndSepPunct{\mcitedefaultmidpunct}
{\mcitedefaultendpunct}{\mcitedefaultseppunct}\relax
\EndOfBibitem
\bibitem[Santos-Cottin \emph{et~al.}(2020)Santos-Cottin, Martino,
  Le~Mardel{\'e}, Witteveen, von Rohr, Homes, Rukelj, and Akrap]{Santos2020}
D.~Santos-Cottin, E.~Martino, F.~Le~Mardel{\'e}, C.~Witteveen, F.~von Rohr,
  C.~Homes, Z.~Rukelj and A.~Akrap, \emph{Physical Review Materials}, 2020,
  \textbf{4}, 021201\relax
\mciteBstWouldAddEndPuncttrue
\mciteSetBstMidEndSepPunct{\mcitedefaultmidpunct}
{\mcitedefaultendpunct}{\mcitedefaultseppunct}\relax
\EndOfBibitem
\bibitem[Choyke \emph{et~al.}(2013)Choyke, Matsunami, and Pensl]{Choyke2013}
W.~J. Choyke, H.~Matsunami and G.~Pensl, \emph{Silicon carbide: recent major
  advances}, Springer Science \& Business Media, 2013\relax
\mciteBstWouldAddEndPuncttrue
\mciteSetBstMidEndSepPunct{\mcitedefaultmidpunct}
{\mcitedefaultendpunct}{\mcitedefaultseppunct}\relax
\EndOfBibitem
\bibitem[Naik and Rastogi(2011)]{Naik2011}
I.~Naik and A.~Rastogi, \emph{Pramana}, 2011, \textbf{76}, 957--963\relax
\mciteBstWouldAddEndPuncttrue
\mciteSetBstMidEndSepPunct{\mcitedefaultmidpunct}
{\mcitedefaultendpunct}{\mcitedefaultseppunct}\relax
\EndOfBibitem
\bibitem[Leroux \emph{et~al.}(2012)Leroux, Le~Tacon, Calandra, Cario, Measson,
  Diener, Borrissenko, Bosak, and Rodiere]{Leroux2012}
M.~Leroux, M.~Le~Tacon, M.~Calandra, L.~Cario, M.~A. Measson, P.~Diener,
  E.~Borrissenko, A.~Bosak and P.~Rodiere, \emph{Physical Review B}, 2012,
  \textbf{86}, 155125\relax
\mciteBstWouldAddEndPuncttrue
\mciteSetBstMidEndSepPunct{\mcitedefaultmidpunct}
{\mcitedefaultendpunct}{\mcitedefaultseppunct}\relax
\EndOfBibitem
\bibitem[Leroux \emph{et~al.}(2018)Leroux, Cario, Bosak, and
  Rodiere]{Leroux2018}
M.~Leroux, L.~Cario, A.~Bosak and P.~Rodiere, \emph{Physical Review B}, 2018,
  \textbf{97}, 195140\relax
\mciteBstWouldAddEndPuncttrue
\mciteSetBstMidEndSepPunct{\mcitedefaultmidpunct}
{\mcitedefaultendpunct}{\mcitedefaultseppunct}\relax
\EndOfBibitem
\bibitem[Carmalt \emph{et~al.}(2004)Carmalt, Manning, Parkin, Peters, and
  Hector]{Carmalt2004}
C.~J. Carmalt, T.~D. Manning, I.~P. Parkin, E.~S. Peters and A.~L. Hector,
  \emph{J. Mater. Chem.}, 2004, \textbf{14}, 290\relax
\mciteBstWouldAddEndPuncttrue
\mciteSetBstMidEndSepPunct{\mcitedefaultmidpunct}
{\mcitedefaultendpunct}{\mcitedefaultseppunct}\relax
\EndOfBibitem
\bibitem[Naito and Tanaka(1982)]{Naito1982}
M.~Naito and S.~Tanaka, \emph{Journal of the Physical Society of Japan}, 1982,
  \textbf{51}, 219--227\relax
\mciteBstWouldAddEndPuncttrue
\mciteSetBstMidEndSepPunct{\mcitedefaultmidpunct}
{\mcitedefaultendpunct}{\mcitedefaultseppunct}\relax
\EndOfBibitem
\bibitem[Onabe \emph{et~al.}(1978)Onabe, Naito, and Tanaka]{Onabe1978}
K.~Onabe, M.~Naito and S.~Tanaka, \emph{Journal of the Physical Society of
  Japan}, 1978, \textbf{45}, 50--58\relax
\mciteBstWouldAddEndPuncttrue
\mciteSetBstMidEndSepPunct{\mcitedefaultmidpunct}
{\mcitedefaultendpunct}{\mcitedefaultseppunct}\relax
\EndOfBibitem
\bibitem[Guillam{\'{o}}n \emph{et~al.}(2008)Guillam{\'{o}}n, Suderow, Vieira,
  Cario, Diener, and Rodi{\`{e}}re]{Guillamon2008a}
I.~Guillam{\'{o}}n, H.~Suderow, S.~Vieira, L.~Cario, P.~Diener and
  P.~Rodi{\`{e}}re, \emph{Phys. Rev. Lett.}, 2008, \textbf{101}, 166407\relax
\mciteBstWouldAddEndPuncttrue
\mciteSetBstMidEndSepPunct{\mcitedefaultmidpunct}
{\mcitedefaultendpunct}{\mcitedefaultseppunct}\relax
\EndOfBibitem
\bibitem[{Van Maaren} and Schaeffer(1966)]{VanMaaren1966}
M.~{Van Maaren} and G.~Schaeffer, \emph{Physics Letters}, 1966, \textbf{20},
  131\relax
\mciteBstWouldAddEndPuncttrue
\mciteSetBstMidEndSepPunct{\mcitedefaultmidpunct}
{\mcitedefaultendpunct}{\mcitedefaultseppunct}\relax
\EndOfBibitem
\bibitem[Ka{\v{c}}mar{\v{c}}{\'\i}k
  \emph{et~al.}(2010)Ka{\v{c}}mar{\v{c}}{\'\i}k, Pribulov{\'a}, Marcenat,
  Klein, Rodi{\`e}re, Cario, and Samuely]{Kacmarcik2010}
J.~Ka{\v{c}}mar{\v{c}}{\'\i}k, Z.~Pribulov{\'a}, C.~Marcenat, T.~Klein,
  P.~Rodi{\`e}re, L.~Cario and P.~Samuely, \emph{Physical Review B}, 2010,
  \textbf{82}, 014518\relax
\mciteBstWouldAddEndPuncttrue
\mciteSetBstMidEndSepPunct{\mcitedefaultmidpunct}
{\mcitedefaultendpunct}{\mcitedefaultseppunct}\relax
\EndOfBibitem
\bibitem[Ka{\v{c}}mar{\v{c}}{\'\i}k
  \emph{et~al.}(2010)Ka{\v{c}}mar{\v{c}}{\'\i}k, Pribulov{\'a}, Marcenat,
  Klein, Rodi{\`e}re, Cario, and Samuely]{Kacmarcik2010b}
J.~Ka{\v{c}}mar{\v{c}}{\'\i}k, Z.~Pribulov{\'a}, C.~Marcenat, T.~Klein,
  P.~Rodi{\`e}re, L.~Cario and P.~Samuely, \emph{Physica C: Superconductivity
  and its applications}, 2010, \textbf{470}, S719--S720\relax
\mciteBstWouldAddEndPuncttrue
\mciteSetBstMidEndSepPunct{\mcitedefaultmidpunct}
{\mcitedefaultendpunct}{\mcitedefaultseppunct}\relax
\EndOfBibitem
\bibitem[Fisher and Sienko(1980)]{Fisher1980}
W.~G. Fisher and M.~Sienko, \emph{Inorganic Chemistry}, 1980, \textbf{19},
  39--43\relax
\mciteBstWouldAddEndPuncttrue
\mciteSetBstMidEndSepPunct{\mcitedefaultmidpunct}
{\mcitedefaultendpunct}{\mcitedefaultseppunct}\relax
\EndOfBibitem
\bibitem[Bardeen \emph{et~al.}(1957)Bardeen, Cooper, and
  Schrieffer]{Bardeen1957}
J.~Bardeen, L.~N. Cooper and J.~R. Schrieffer, \emph{Phys. Rev.}, 1957,
  \textbf{108}, 1175--1204\relax
\mciteBstWouldAddEndPuncttrue
\mciteSetBstMidEndSepPunct{\mcitedefaultmidpunct}
{\mcitedefaultendpunct}{\mcitedefaultseppunct}\relax
\EndOfBibitem
\bibitem[Carnicom \emph{et~al.}(2019)Carnicom, Xie, Yang, G{\'o}rnicka, Kong,
  Klimczuk, and Cava]{Carnicom2019}
E.~M. Carnicom, W.~Xie, Z.~Yang, K.~G{\'o}rnicka, T.~Kong, T.~Klimczuk and
  R.~J. Cava, \emph{Chemistry of Materials}, 2019, \textbf{31},
  2164--2173\relax
\mciteBstWouldAddEndPuncttrue
\mciteSetBstMidEndSepPunct{\mcitedefaultmidpunct}
{\mcitedefaultendpunct}{\mcitedefaultseppunct}\relax
\EndOfBibitem
\bibitem[von Rohr \emph{et~al.}(2014)von Rohr, Luo, Ni, W{\"o}rle, and
  Cava]{VonRohr2014}
F.~von Rohr, H.~Luo, N.~Ni, M.~W{\"o}rle and R.~J. Cava, \emph{Physical Review
  B}, 2014, \textbf{89}, 224504\relax
\mciteBstWouldAddEndPuncttrue
\mciteSetBstMidEndSepPunct{\mcitedefaultmidpunct}
{\mcitedefaultendpunct}{\mcitedefaultseppunct}\relax
\EndOfBibitem
\bibitem[Leroux(2012)]{Leroux2012b}
M.~Leroux, \emph{PhD thesis}, 2012\relax
\mciteBstWouldAddEndPuncttrue
\mciteSetBstMidEndSepPunct{\mcitedefaultmidpunct}
{\mcitedefaultendpunct}{\mcitedefaultseppunct}\relax
\EndOfBibitem
\bibitem[Katzke(2002)]{Katzke2002}
H.~Katzke, \emph{Zeitschrift f{\"u}r Kristallographie-Crystalline Materials},
  2002, \textbf{217}, 127--130\relax
\mciteBstWouldAddEndPuncttrue
\mciteSetBstMidEndSepPunct{\mcitedefaultmidpunct}
{\mcitedefaultendpunct}{\mcitedefaultseppunct}\relax
\EndOfBibitem
\end{mcitethebibliography}
\bibliographystyle{rsc} 

\end{document}